\documentclass[twocolumn,showpacs,preprintnumbers,amsmath,amssymb]{revtex4}

\usepackage{graphicx}
\usepackage{dcolumn}
\usepackage{bm}
\usepackage{color}

\begin{document}

\preprint{APS/123-QED}

\title{Magnetic correlations in the pressure-induced superconductor CrAs investigated by $^{75}$As nuclear magnetic resonance}

\author{Kei Matsushima$^1$, Hisashi Kotegawa$^{1}$, Yoshiki Kuwata$^1$, Hideki Tou$^1$, Jun Kaneyoshi$^1$, \\
Eiichi Matsuoka$^1$, Hitoshi Sugawara$^1$, Takahiro Sakurai$^{2}$, Hitoshi Ohta$^{1,3}$, and Hisatomo Harima$^1$}

\affiliation{
$^1$Department of Physics, Kobe University, Kobe 657-8501, Japan \\
$^{2}$Research Facility Center for Science and Technology, Kobe University, Kobe, Hyogo 657-8501, Japan\\
$^{3}$Molecular Photoscience Research Center, Kobe University, Kobe, Hyogo 657-8501, Japan
}

\date{\today}

\begin{abstract}
We report $^{75}$As-NMR results for CrAs under pressure, which shows superconductivity adjoining a helimagnetically ordered state.
We successfully evaluated the Knight shift from the spectrum, which is strongly affected by the quadrupole interaction. 
The Knight shift shows the remarkable feature that the uniform spin susceptibility increases toward low temperatures in the paramagnetic state.
This is in sharp contrast to CrAs at ambient pressure, and also to cuprates and Fe pnictides, where antiferromagnetic correlations are dominant.
Superconductivity emerges in CrAs under unique magnetic correlations, which probably originate in the three-dimensional zigzag structure of its nonsymmorphic symmetry.

\end{abstract}

\maketitle

The interplay between magnetism and superconductivity has long been an attractive subject in condensed matter physics.
If magnetism favors the occurrence of superconductivity, it can yield unconventional superconducting (SC) pairing without requiring mediation by electron-phonon coupling.
Simultaneously, it induces a wide variety of features in the SC state, which is affected by many factors, such as the kinds of magnetic interactions, the topology of the Fermi surface, the presence/absence of time-reversal symmetry, {\it etc.}
Chromium arsenide (CrAs) is a rare Cr-based example that is thought to be in this category, and it shows a pressure-induced quantum phase transition from a helimagnetically ordered state to a SC state.\cite{Wu,Kotegawa}
In spite of the strong first-order character of the magnetic phase transition,\cite{Boller,Suzuki,Kotegawa_NQR,Khasanov} the development of magnetic fluctuations in the paramagnetic (PM) state has been observed under pressure by the nuclear spin-lattice relaxation rate $1/T_1$,\cite{Kotegawa_NQR} and by recent inelastic neutron scattering measurements for the high-temperature PM phase in polycrystalline CrAs and for P-doped CrAs.\cite{Matsuda} 
The presence of strong electronic correlations is also supported by the non-Fermi-liquid like behavior of the resistivity \cite{Wu,Kotegawa,Matsuda} and the lack of observations of quantum oscillations.\cite{Niu} 
Our previous nuclear quadrupole resonance (NQR) study suggests that these fluctuations have no direct association with the first-order magnetic transition, and we speculate that it may originate from a proximity to a hidden instability.\cite{Kotegawa_JP}
This point is still open to argument, but the absence of a Hebel-Slichter peak in $1/T_1$ and the close relationship between the magnetic fluctuations and the superconductivity suggest the possibility that the superconductivity of CrAs is mediated by magnetic interactions beyond the BCS framework.\cite{Kotegawa_NQR} 
The details of the magnetic fluctuations are important for understanding superconductivity in CrAs, but they are still unclear, especially for pure CrAs under pressure.
If the quantum magnetic phase transition is of second order, we can infer the wave vector of the magnetic fluctuations from the magnetic structure of the ordered state.
In CrAs, however, this connection is not guaranteed because of the significant segregation of the electronic states in the two phases due to the strong first-order phase transition. 
A direct measurement in the PM state is therefore crucial for understanding the magnetic correlations that have possible connections with the superconductivity.

Another notable feature of CrAs is its crystal structure, which belongs to the nonsymmorphic space group $Pnma$.
The nearest-neighbor Cr ions form a zigzag chain along the $a$-axis with a bond length of $\sim2.89$ \AA, while the second- and third-nearest-neighbor Cr ions at $\sim3.03$ \AA \ and $\sim3.46$ \AA, respectively, form a zigzag ladder along the $b$ axis.
In this isostructural three-dimensional zigzag structure, helimagnetic phases often appear, as observed in MnP, FeP, and FeAs.\cite{Forsyth,Felcher2,Selte}
It is thought that competition among the exchange interactions between the multiple neighboring magnetic ions yields the characteristic helimagnetic phases.\cite{Takeuchi,Kallel,Dobrzynski} 
The structure of CrAs differs completely from the typical layered structure such as cuprates and Fe-pnictides.
It belongs to the same space group as those of the pressure-induced superconductor MnP \cite{Cheng} and U-based ferromagnetic (FM) superconductors.\cite{Aoki,Huy}
It is an intriguing subject to clarify what types of magnetic fluctuations develop in this zigzag structure in CrAs, how they affect the superconductivity, and what kind of SC symmetry is realized as a consequence. 
In this Rapid Communication, we present nuclear magnetic resonance (NMR) studies under pressure to evaluate the magnetic correlations in CrAs.


The single crystals of CrAs were prepared using the Sn-flux method from a starting composition of ${\rm Cr}:{\rm As}:{\rm Sn} = 1:1:20$ avoiding the structural phase transition during crystal growth.\cite{Kotegawa_JP}
For the NMR measurements under pressure, we generate hydrostatic pressure using an indenter type of pressure cell and Daphne 7474 as a pressure transmitting medium.\cite{indenter,Murata}
Pressure calibration was performed using the superconducting transition temperature of a lead manometer.
NMR measurements were performed utilizing As nuclei with a nuclear spin $I = 3/2$ and a gyromagnetic ratio $\gamma = 7.292$ MHz/T.
A single crystal of about 1 mm in size was used for the NMR measurements under pressure.
We adjusted the angle of the single crystal in the pressure cell relative to the magnetic field by utilizing a uniaxial rotator mounted on the NMR probe.
A magnetic susceptibility up to 340 K was measured by utilizing a magnetic property measurement system (MPMS, Quantum Design). 
For the measurements, several single crystals were aligned so that $H \parallel b$.

\begin{figure}[htb]
\centering
\includegraphics[width=0.9\linewidth]{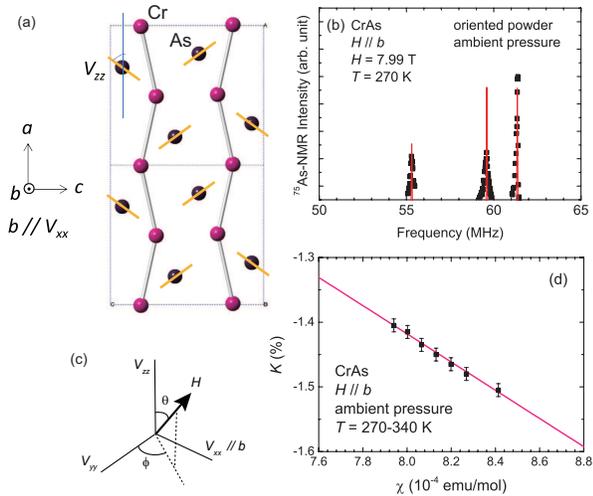}
\caption[]{(color online) (a) The crystal structure of CrAs and the directions of the principal axes of the EFG at the As site. The orange bars indicate the directions of $V_{zz}$. (b) The frequency-swept NMR spectrum in the PM state at ambient pressure. The red curves indicate the simulation for $H \parallel b \parallel V_{xx}$. (c) The polar angles between the magnetic field and the principal axes of the EFG. (d) $K$-$\chi$ plot at ambient pressure. A negative hyperfine coupling of $A = - 12.2 \pm 0.5$ T/$\mu_B$ and $K_{orb}=0.326 \pm 0.07$\% are obtained.
}
\end{figure}

As already reported in our $^{75}$As-NQR study,\cite{Kotegawa_NQR} the As nuclei have a large quadrupole interaction in CrAs, which significantly affects the shape of the NMR spectrum; therefore, an evaluation of the Knight shift requires careful measurements and analyses that consider the electric-field-gradient (EFG) tensor at the nuclear site.\cite{Kotegawa_PhysicaB}
Figure 1(a) shows the crystal structure of CrAs and the directions of the principal axes of the EFG; $V_{zz}$ and $V_{xx}$, at the As sites.
Here $V_{ii}$ ($i = x, y$, and $z$) are the values of the EFG tensor along the principal axis $i$, and $ |V_{zz}|> |V_{yy}|> | V_{xx} | $.
In CrAs, the direction of $V_{zz}$ is obtained from the band structure calculation,\cite{Kotegawa_PhysicaB} as shown in the figure.
It lies in the $ac$ plane and is tilted by 54.5$^{\circ}$ from the $a$-axis; that is, by 35.5$^{\circ}$ from the $c$-axis.
One of the principal axes is restricted to lie along the $b$ axis from the local symmetry of $.m.$ at the As site, and the band structure calculation and previous NMR measurements at ambient pressure demonstrate that it is $V_{xx}$.\cite{Kotegawa_PhysicaB}
Two As sites with different $V_{zz}$ directions arise in the unit cell because of the nonsymmorphic symmetry, but they are crystallographically equivalent.
When a magnetic field is applied precisely along any direction among the crystal axes $a$, $b$ or $c$, all the As sites maintain equivalency.

Figure 1(b) shows the frequency-swept $^{75}$As-NMR spectrum in the PM state at ambient pressure. Here, the powdered crystals were oriented with $H \parallel b$, which is the easy axis,\cite{Wu_single} and we observed three NMR transitions for $I=3/2$.
The NMR spectral shape including the quadrupole interactions is determined by five parameters; the Knight shift, the quadrupole frequency $\nu_Q$, the asymmetric parameter $\eta$, and the polar angles ($\theta$, $\phi$) between $H$ and the principal axes of the EFG tensor, as shown in Fig.~1(c).
For $H \parallel b \parallel V_{xx}$, we can fix $\theta=90^{\circ}$ and $\phi=90^{\circ}$; therefore, the observation of three transitions makes it possible to derive separately the Knight shift, $\nu_Q$, and $\eta$.
The numerical simulation shown by the red curves reproduces the experimental data well.
From the temperature dependence of the Knight shift, we obtained the $K$-$\chi$ plot for $H \parallel b$ in the PM state at ambient pressure, as shown in Fig.~1(d).
Here, we used susceptibility data measured at 1 T for our single crystals.
The Knight shift is generally composed of a temperature-dependent spin part and a temperature-independent orbital part, as follows;
\begin{eqnarray}
K = K_{spin} + K_{orb} = \frac{A}{N\mu_B}\chi_{spin} + K_{orb}
\end{eqnarray}
Here, $N$ is the Avogadro's number, and $\mu_B$ is the Bohr magneton.
The linear relationship in Fig.~1(d) gives the hyperfine coupling constant $A = - 12.2 \pm 0.5$ T/$\mu_B$ and $K_{orb} = 0.326 \pm 0.07 \%$ for $H \parallel b$.
The negative value of $A$ suggests that the core-polarization effect is dominant in the hyperfine coupling, due to the hybridization of the Cr-$3d$ and As-$4p$ orbitals.

\begin{figure}[htb]
\centering
\includegraphics[width=\linewidth]{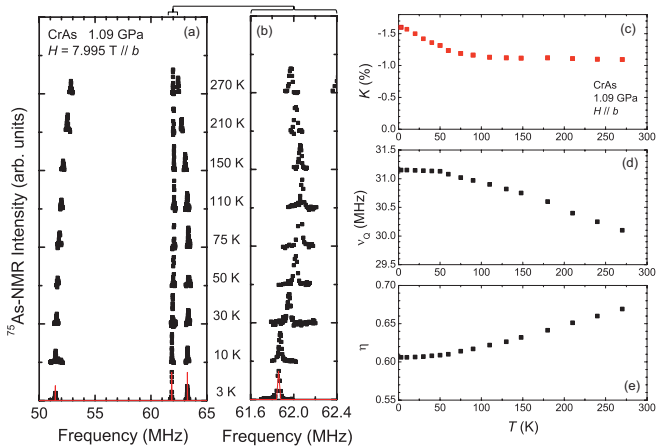}
\caption[]{(color online) (a) $^{75}$As-NMR spectra at 1.09 GPa for various temperatures. At this pressure, the PM state is stable down to the lowest temperature. All the data can be reproduced with $H \parallel b \parallel V_{xx}$ as shown, for example, by the red curve at 3 K. (b) The NMR spectra at around 62 MHz. (c-e) The temperature dependences of the Knight shift, $\nu_Q$, and $\eta$, which are derived separately from the NMR spectra. The Knight shift increases in the negative direction with decreasing temperature, indicating that the spin susceptibility increases toward low temperatures because of the negative hyperfine coupling due to the hybridization of Cr-3$d$ and As-$4p$ electrons. 
}
\end{figure}

Figure 2(a) shows the temperature variation of the $^{75}$As-NMR spectrum for $H \parallel b$ at $1.09$ GPa, where the helimagnetic phase is completely suppressed.
Superconductivity appears below $\sim2$ K, but it is suppressed by an applied field $H \sim 8$ T.
The observation of three clear NMR transitions for $H \parallel b$ ensures the precise orientation of the single crystal in the pressure cell relative to the magnetic field. 
The spectrum can be reproduced well for the condition $H \parallel V_{xx}$, as is also the case at ambient pressure, as shown by the red curves at 3 K.
The temperature dependences of the Knight shift, $\nu_Q$, and $\eta$ are shown in Figs. 2(c-e).
The temperature dependences of $\nu_Q$ and $\eta$ probably originate in the thermal expansion of the lattice parameters.
The estimated values of $\nu_Q$ and $\eta$ at the lowest temperature give the NQR frequency of 33 MHz, which is consistent with the actual resonance frequency at 1.07 GPa,\cite{Kotegawa_NQR} confirming the validity of the present determination.
The vertical axis in Fig.~ 2(c) shows that the Knight shift increases in the negative direction because the hyperfine coupling is negative.
From the relationship $K_{spin} \propto A \chi_{spin}$, we found that $\chi_{spin}$ increases toward low temperatures in the PM state where superconductivity appears.
This is qualitatively consistent with recent bulk-susceptibility measurements under pressure, although the low-temperature data below 50 K are not shown.\cite{Matsuda}
Our microscopic measurement, which is insensitive to extrinsic contributions, shows that the increase of $\chi_{spin}$ toward the lowest temperature is inherent to the PM state of CrAs.

\begin{figure}[htb]
\centering
\includegraphics[width=0.8\linewidth]{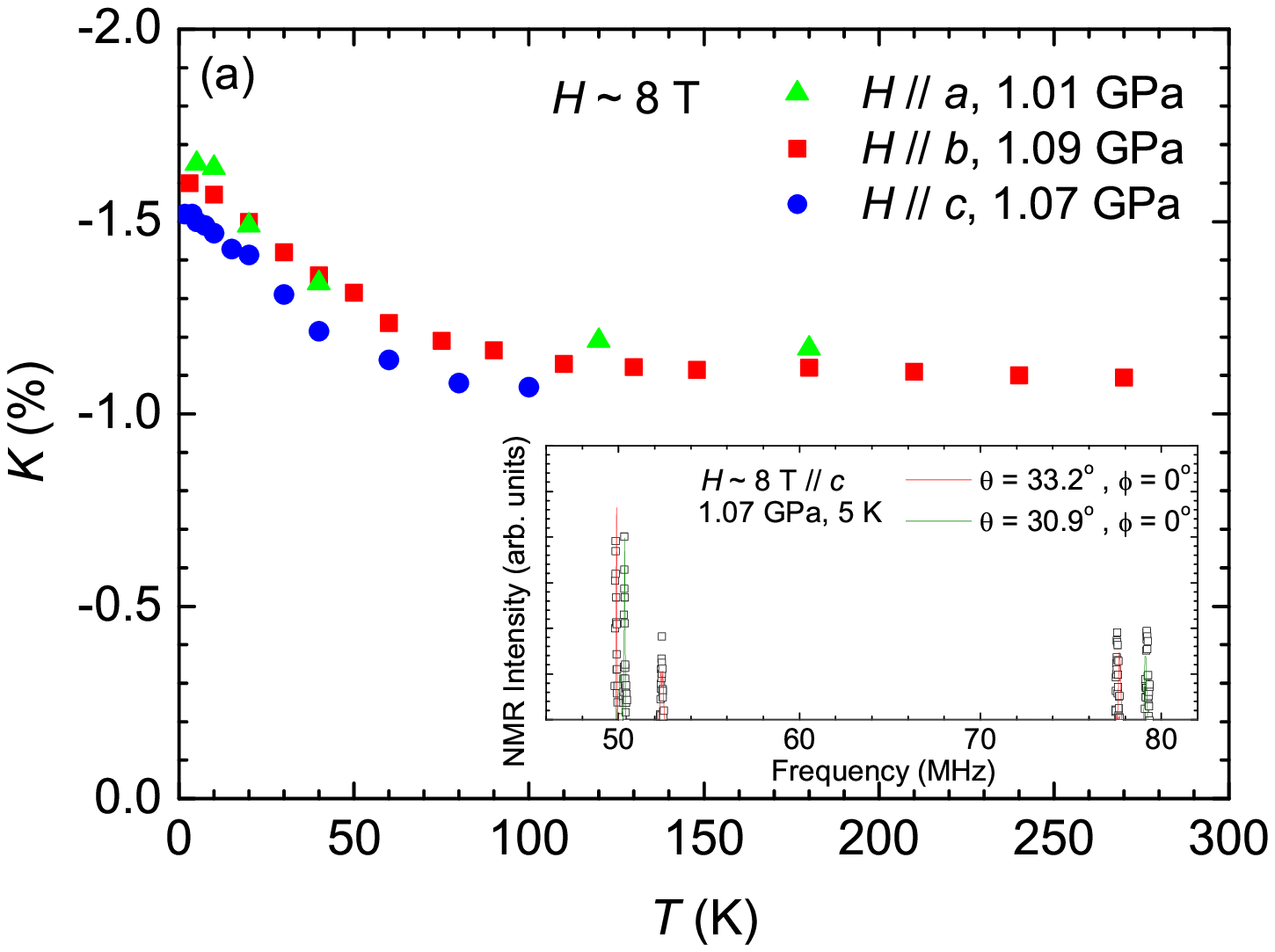}
\includegraphics[width=0.8\linewidth]{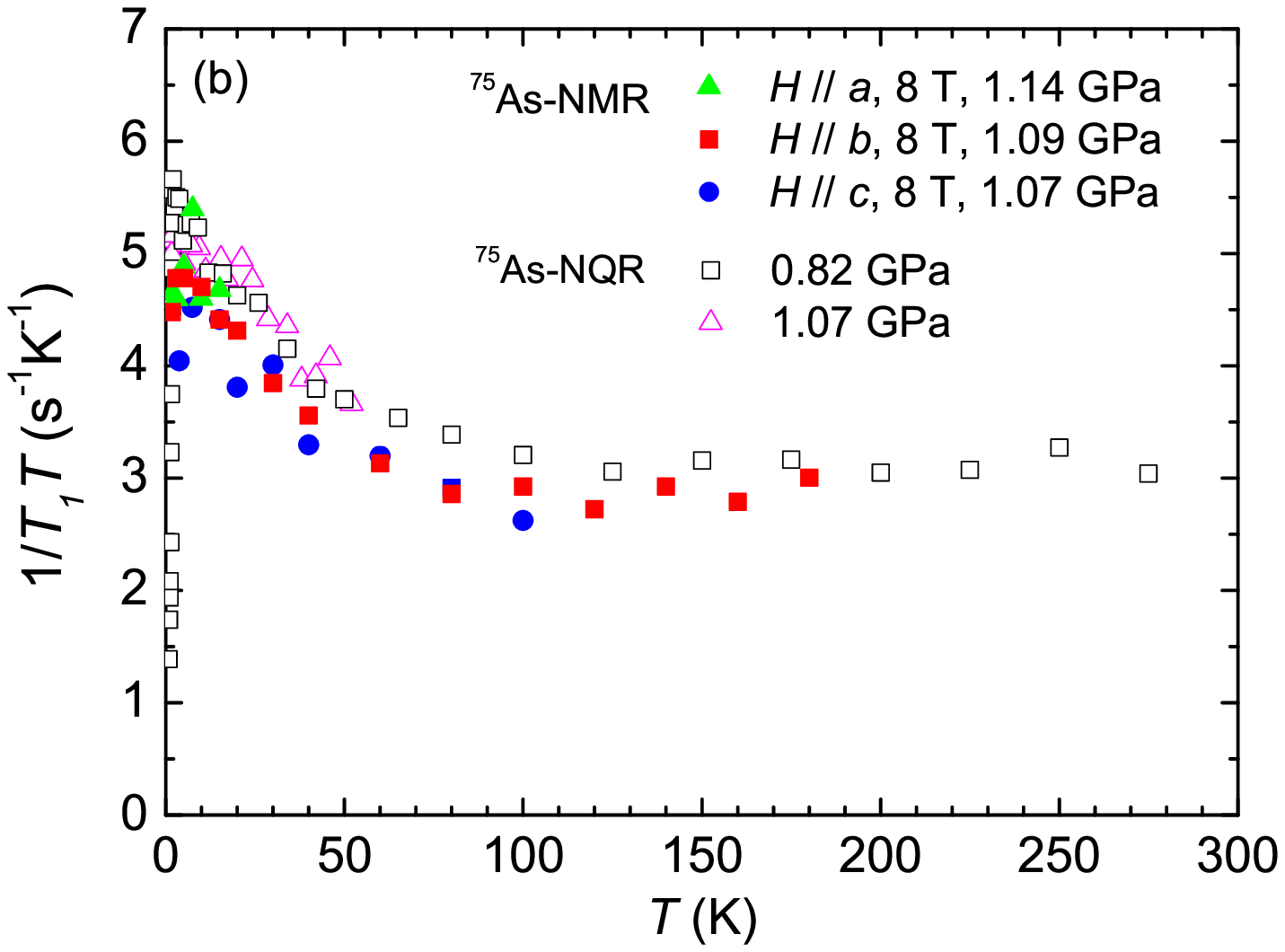}
\caption[]{(color online) The temperature dependences of (a) the Knight shift and (b) $1/T_1T$ under pressure. A magnetic field of $\sim8$ T is applied separately along each of the three crystal axes. Neither the Knight shift nor $1/T_1T$ show any significant anisotropy. 
}
\end{figure}

At ambient pressure, a magnetic anisotropy of about $10-20$ \% has been observed in the bulk susceptibility in CrAs.\cite{Wu_single,Zhu}
This may imply the presence of a substantial spin-orbit interaction, and the intense lattice change in the first-order magnetic transition also implies this possibility.\cite{Boller,Suzuki}
On the other hand, the spin-orbit interaction has been suggested to contribute to generating the ellipticity of the spin helix in FeAs, while it seems to be negligible in the ordered state of CrAs.\cite{Rodriguez,Frawley}
To investigate the magnetic anisotropy of CrAs under pressure, we measured the anisotropy of the Knight shift and $1/T_1$.
To determine the Knight shift for $H \parallel a$ and $c$, $\phi$ can be $0^{\circ}$, but $\theta$ cannot be fixed because the $a$ and $c$ axes are not directed along the principal axes of the EFG tensor.
Here, we used the values of $\nu_Q$ and $\eta$ obtained in the measurement for $H \parallel b$ to reduce the number of parameters.
The inset of Fig.~3(a) shows the NMR spectrum for $H\parallel c$ at 1.07 GPa.
The three transitions are split, owing to a misalignment of the single crystal relative to the magnetic field, but all the peaks can be explained by two sets of alignments; ($\theta=33.2^{\circ}$ and $\phi=0^{\circ}$) and ($\theta=30.9^{\circ}$ and $\phi=0^{\circ}$).
As seen in Fig.~1(a), the misalignment yields two inequivalent As sites.
The two sets of $\theta$ indicate that the misalignment is about $(33.2^{\circ}-30.9^{\circ})/2 \simeq 1$$^{\circ}$.
The angle between $V_{zz}$ and the $c$ axis is estimated to be $(33.2^{\circ}+30.9^{\circ})/2 = 32.05^{\circ}$, which agrees with the value 35.5$^{\circ}$ obtained from band structure calculations using the crystal structure at ambient pressure.\cite{Kotegawa_NQR} 
The resulting temperature dependence of the Knight shift is shown in Fig. 3(a).
The Knight shifts show similar values and temperature dependences for all three axes.
From the NMR spectrum for the powdered sample,\cite{Kotegawa_PhysicaB} the hyperfine coupling constant is expected to be almost isotropic, which seems reasonable because the core-polarization effect dominates the coupling; therefore, we expect the anisotropy of the magnetic susceptibility of CrAs to be not strong under pressure. 
Figure 3(b) shows the temperature dependence of $1/T_1T$ under a magnetic field together with the NQR measurement at zero field.\cite{Kotegawa_NQR}
The developments of $1/T_1T$ toward low temperatures are all similar and the anisotropy is not significant. 
The $1/T_1$ generally corresponds to magnetic fluctuations perpendicular to the nuclear quantization axis.
For the NQR measurement at zero field, the nuclear quantization axis is determined by the EFG tensor.
Because of the large asymmetric parameter $\eta$ at the As site, the main contribution to $1/T_1$ in NQR is from the magnetic fluctuations along $V_{xx}$, that is along the $b$ axis.
For $I=3/2$ and $\eta=0.61$, for instance, the magnetic fluctuations along the $b$ axis constitute $\sim64$\% of the magnetic fluctuations that contribute to $T_1$.\cite{Chepin}
If the magnetic field is much greater than the quadrupole interaction, the quantization axis is directed along the magnetic field.
For the As nuclei, a magnetic field of 8 T corresponds to $\gamma H \sim 58.3$ MHz, which is about twice the NQR frequency.
This does not satisfy the condition that the Zeeman interaction is dominant over the quadrupole interaction, but $1/T_1$ for each axis is expected to contain sufficient components perpendicular to the magnetic field. 
Therefore, the similarity in $1/T_1T$ among the different magnetic field orientations suggests that the anisotropy of the magnetic fluctuations is not strong in CrAs.

\begin{figure}[htb]
\centering
\includegraphics[width=\linewidth]{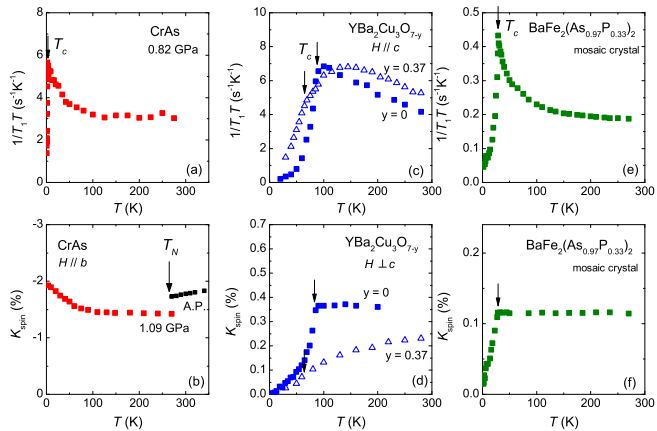}
\caption[]{(color online) Comparison of $1/T_1T$ and $K_{spin}$ among CrAs, YBa$_2$Cu$_3$O$_{7-y}$, and BaFe$_2$(As$_{1-x}$P$_x$)$_2$. The data for YBa$_2$Cu$_3$O$_{7-y}$ and BaFe$_2$(As$_{1-x}$P$_x$)$_2$ are taken from Refs.~\cite{Takigawa1,Takigawa2, Nakai}. In panel(b), the data at ambient pressure (A.P.) are also shown, and superconductivity at 1.09 GPa is suppressed by the magnetic field.
}
\end{figure}

The magnetic correlations of CrAs under pressure exhibit the feature that the spin susceptibility increases toward low temperatures, in sharp contrast to other typical $d$-electron superconductors such as cuprates and Fe-pnictides.\cite{Takigawa1,Takigawa2,Nakai,Ning,Imai}
Figure 4 shows comparisons of $1/T_1T$ and $K_{spin}$ for YBa$_2$Cu$_3$O$_{7-y}$ and BaFe$_2$(As$_{1-x}$P$_x$)$_2$.\cite{Takigawa1,Takigawa2,Nakai}
Here, we assume that $K_{orb}=0.326$ \%, independent of pressure, in order to deduce $K_{spin}$ for CrAs.
All three systems show a Curie-Weiss like increase in $1/T_1T$, indicative of the development of low-energy magnetic excitations.
The $K_{spin}$ in optimally doped YBa$_2$Cu$_3$O$_{7-y}$ with $y=0$ is independent of temperature, and it decreases in the underdoped region with $y=0.37$.
In BaFe$_2$(As$_{1-x}$P$_x$)$_2$ with $x=0.33$, $K_{spin}$ is temperature independent.
The $1/T_1T$ is proportional to the $q$-summed imaginary part of the dynamical susceptibility, as shown by $1/T_1T \propto \Sigma_q \chi''(q, \omega_N)/\omega_N$, where $\omega_N$ is the resonance frequency, while the Knight shift is proportional to $\chi(q=0,\omega=0)$. 
In cuprates and Fe-pnictides, it is thought that the spectral weight near $q=0$ is not affected or is even suppressed by the development of antiferromagnetic (AF) fluctuations. 
In itinerant AF systems, the behavior of the spin susceptibility in the PM state depends on the material, and an almost-temperature-independent susceptibility is also observed in the typical spin density wave system Cr.\cite{Fawcett}
In CrAs, which must be in the itinerant regime, the susceptibility at ambient pressure does not follow a Curie-Weiss behavior and is significantly suppressed toward $T_N$.\cite{Saparov}
The Knight shift below 340 K at ambient pressure is shown in Fig.~4(b).
The suppression of the susceptibility seems to suggest that uniform susceptibility is not important for the magnetic correlations in CrAs; however, our result under pressure shows that a uniform susceptibility, that is ferromagnetic-like correlations, develops significantly toward the emergence of superconductivity. 
If the ferromagnetic correlations are dominant in a system, $K_{spin}$ should develop in proportion to $1/T_1T$ following $1/T_1T \propto K_{spin}^n$ with $n=1$, as predicted from self-consistent renormalization theory.\cite{Moriya} 
The experimental data for CrAs follow the relationship $1/T_1T \propto K_{spin}^n$ with $n=1.77 \pm 0.07$, which differs from the typical FM case.
This result indicates that the dominant $q$ vector is finite in CrAs, but the comparable increase of $1/T_1T$ and $K_{spin}$ suggests that magnetic correlations near $q=0$ are also significant.
This interpretation is not contradictory to recent inelastic neutron scattering measurements at ambient pressure, which suggest a spin-wave excitation with a finite $q$ vector in the PM state, but this remains uninvestigated near $q=0$.\cite{Matsuda}
Our results suggest that the magnetic correlations in CrAs differ from a case in which the spectral weight only develops at a finite $q$ vector.
This is likely to be related to the competition among the exchange interactions with different signs that originate in the structural features.\cite{Takeuchi,Kallel,Dobrzynski}
In isostructural MnP, this character is evidenced by the multiple magnetic phases generated by temperature and pressure.\cite{Cheng,Matsuda2} 
The NMR results for CrAs under pressure are similar to those of another Cr-based superconductor $A$$_2$Cr$_3$As$_3$ ($A=$ K, Rb, Cs), where both $1/T_1T$ and $K_{spin}$ increase toward low temperatures.\cite{Yang,Zhi}
This system is composed of Cr-triangular lattices, and CrAs is the distorted triangular lattice.\cite{Kotegawa_JP}
It may thus be interesting to investigate possible underlying similarities between these two systems. 
Another interesting aspect of CrAs is the possibility of spin-triplet superconductivity, which has been suggested from the anisotropy of the upper critical field.\cite{Guo}
It will be intriguing to investigate the interplay between the unique magnetic correlations in CrAs and superconductivity.
The Knight shift measurements in the SC state are crucial to determining the parity of SC pairing in CrAs, and such experiments are currently being planned.

In conclusion, we obtained several features of the magnetic correlations in CrAs through $^{75}$As-NMR measurements.
Measurements of the Knight shift and $1/T_1$ under different magnetic -field directions reveal that magnetic anisotropy is not significant in the PM state of CrAs.
The Knight shift in the normal state suggests that a uniform spin susceptibility develops significantly toward the emergence of superconductivity.
This may originate in the crystal structure, which consists of several kinds of neighboring Cr-Cr bonds, and it will be a key ingredient for understanding superconductivity in CrAs.

\section*{Acknowledgements}

We thank Masaaki Matsuda and Tetsuro Kubo for helpful discussions.
This work was supported by JSPS KAKENHI Grant Number JP15H05882,
JP15H05885, and JP18H04321 (J-Physics), 15H03689.and 15H05745.

\end{document}